# Modeling kinetics of the natural aggregation structures and processes with the solution of generalized logistic equation

Lev A Maslov[1], Vladimir I Chebotarev[2]

[1]University of Northern Colorado, Aims College, Greeley, CO, USA,

lev.maslov@aims.edu

[2]Computer Center RAS, Khabarovsk, Russia, chebotarev@as.khb.ru

**Abstract**

The generalized logistic equation is derived to model kinetics and statistics of natural processes such as earthquakes, forest fires, floods, landslides, and many others. This equation has the form

$$\frac{dN(A)}{dA} = s \cdot (1 - N(A)) \cdot N(A)^q \cdot A^{-\alpha},$$

$q>0$ and $A>0$ is the size of an element of a structure, and $\alpha \geq 0$. The function $N(A)$ can be understood as an approximation to the number of elements of size $A$. The general solution of this equation for $q=1$ is a product of an increasing bounded function and power-law function with stretched exponential cut-off; the power-law distribution is asymptotically nested in the stretched exponential distribution. The relation with Tsallis non-extensive statistics is demonstrated by solving the generalized logistic equation for $q>0$. In the case $0<q<1$ the equation models super-additive, and the case $q>1$ it models sub-additive structures. The Gutenberg-Richter (G-R) formula results from interpretation of empirical data as a straight line in the area of stretched exponent with small $\alpha$. The solution is applied for modeling distribution of foreshocks and aftershocks in the regions of Napa Valley 2014, and Sumatra 2004 earthquakes fitting the observed data well, both qualitatively and quantitatively.

## Introduction

A probability distribution function with a power-law dependence in terms of a process or object sizes is the very common statistical feature of many natural processes and systems (Clauset, et al., 2009; Sornette, 2009). To account quantitatively for fat tail distributions, J. Laherr`ere and D. Sornette (1998), proposed the combination of power-law function with stretched exponential cut-off. Statistical studies (Clauset, et al., 2009) show that in some cases the power law is ruled out, but in the other cases, conjectures are consistent with a power law-distribution. However, once the evidence for a power law distribution is demonstrated, the uncertainty – what physical process is behind this distribution – remains. A proper solution of kinetic equation modeling the process should be complementary to the results of statistical research. Examples of this approach are fracture kinetics equations developed to explain clustering of earthquake sequences (Newman, Knopoff, 1983) and fracture kinetics equations developed by Z. Czechowski (1991). The authors of (Ovchinnikov, Eliseev, 1981) use the logistic equation to describe the process of corrosion fracture registering signals of acoustic emission. J. D. Achter and C. T. Webb (2006) and J. Goldenberg with coauthors (2000) use logistic equation to describe aggregation processes in biology and in marketing. F. Vallianatos and coauthors (Vallianatos, Michas, Papadakis, 2015) presented kinetic equation to model evolution of structures in the frame of non-extensive physics statistics. However, the main lack of equations mentioned above is that they do not have solutions with the fractional power-law exponent. Publications (Gabrielov, et al., 1999; Yakovlev, et.al., 2005) demonstrate the other approach to modeling the "power-law" structures and processes. The authors developed the formal model of hierarchical aggregation in the inverse cascade process. In this model, the aggregation of elements, similar to that of aggregation of clusters in percolation, produces a fractal hierarchical structure. These authors showed that the distribution of elements in their model approaches asymptotically the power-law distribution with fractional exponent $\alpha$:

$$N_i \propto A_i^{-\alpha}, \ \ \alpha \ge 0, \ i >> 1.$$

In the current work we: a) combine two approaches to derive kinetic equation to model hierarchical aggregation structures and processes, b) analyze solution of this equation and to show its relation with power-law function with stretched exponential cut-off, and c) demonstrate relation of general solution with Tsallis non-extensive statistics.

**Kinetic equation for $q = 1$**

Let $N(A)$ is a differentiable function of a real variable $A \ge 0$. This variable represents the "size" of an element in a structure (it can be an area, mass, magnitude of earthquake, etc.). The function $N(A)$ can be understood as an approximation to the number of all elements with the size less than $A$. Considering $A$ as a "natural variable" similar to that used in thermodynamics, we suggest that the rate of growth of $N$ with $A$ is proportional to product $N(A) \cdot A^{-\alpha}$ :

$$\left. \frac{dN(A)}{dA} \right|_{gain} = s \cdot N(A) \cdot A^{-\alpha}, \ q{=}1, \ s{>}0, \tag{1}$$

where $A$ is the size of an element of a structure defined by its characteristic diameter and $0 \le \alpha \le 1$. $A^{-\alpha}$ represents the "fractal correction" to the area of an element of a structure. The rate of loss of elements due to their coalescence is

$$\left.\frac{dN(A)}{dA}\right|_{loss} = -L(N) \cdot N(A) \cdot A^{-\alpha} \qquad (2)$$

Function $L(N)$ can be derived analytically for some relatively simple models of aggregation. In this work, we suggest that $L(N) = k \cdot N$, $k$ is a positive constant.

Thus, the balance equation is:

$$\frac{dN(A)}{dA} = [s - k \cdot N(A)] \cdot N(A) \cdot A^{-\alpha} \qquad (3)$$

Normalizing $N$ by $s/k$, s > 0, k > 0, we arrive at the equation

$$\frac{d\bar{N}}{dA} = s \cdot (1 - \bar{N}) \cdot \bar{N} \cdot A^{-\alpha} \quad , \qquad (4)$$

with $\bar{N} = N \cdot k / s$, $\bar{N} < 1$, and $\bar{N} \to 1$ for $A \to \infty$, $\bar{N}(0) = 0$, i.e. $\bar{N}(A)$ is the cumulative distribution function of the random size of an element of a structure. Equation (4) is presented first in (Maslov, Anokhin, 2012). If $0 \leq \alpha < 1$, then equation (4) can be written in the form of a classical logistic equation:

$$\frac{d\tilde{N}}{dx} = s' \cdot (1 - \tilde{N}) \cdot \tilde{N} \qquad (5)$$

with $\tilde{N}(x) = \bar{N}(x^{1/(1-\alpha)})$, $x = A^{1-\alpha}$ and $s' = s/(1-\alpha)$. If $\alpha = 1$ equation (4) can be written in the form (5) with $\tilde{N}(x) = \bar{N}(e^x)$, $x = \ln A$, and $s' = s$.

In the rest part of the work we omit the bar sign over $N$ in (4). The boundary condition for (4) is

$$N(0+) \equiv \lim_{A \to 0} N(A) = N_0 \quad . \qquad (6)$$

The solution to the problem (4), (6) for $0 \leq \alpha < 1$ is:

$$\begin{cases} N(A) = \left(1 + C\exp\left(-\dfrac{s}{1-\alpha}A^{1-\alpha}\right)\right)^{-1}, 0 \le A < \infty \\ N(0+) = N_0 \end{cases} \qquad (7)$$

with the derivative

$$\frac{dN}{dA} = C \cdot s \cdot \frac{e^{-\frac{s}{1-\alpha}A^{1-\alpha}}}{\left(1 + Ce^{-\frac{s}{1-\alpha}A^{1-\alpha}}\right)^2} A^{-\alpha}, \quad A>0. \qquad (8)$$

Function (8) is not a density function because

$$\int_{0+}^{\infty} \frac{dN(A)}{dA}\,dA = \int_{0+}^{\infty} dN(A) = N(\infty) - N(0+) = 1 - N_0 < 1, \qquad (9)$$

but for sufficiently large $C$, $\dfrac{dN}{dA}$ can be considered approximately as a probability density

function on half-axis $A \ge 0$. Note that the function

$$F(A) = \begin{cases} 0, & A < 0, \\ N(A), & A \ge 0 \end{cases}$$

is the cumulative distribution function with the jump $N_0$ at $A = 0$.

For $\alpha = 0$ in (7) we have

$$\begin{cases} N(A) = (1 + C \cdot e^{-sA})^{-1}, 0 \le A < \infty, \\ N(0+) = N_0 \end{cases} \qquad (10)$$

with $C = \dfrac{1 - N_0}{N_0}$. Note that $(1 + C \cdot e^{-sA})^{-1}$ is a distribution function of a logistic law if to consider

it on the axis $(-\infty, \infty)$. For $\alpha = 1$ the solution of (4) is

$$\begin{cases} N(A) = (1 + C \cdot A^{-s})^{-1}, 0 < A < \infty, \\ N(b) = N_b, b > 0 \end{cases} \tag{11}$$

which is the log-logistic distribution function with $C = b^s \dfrac{1 - N_b}{N_b}$.

The complementary to *N(A)* function in (7) $N^* = 1 - N$:

$$\begin{cases} N^*(A) = \dfrac{C \cdot e^{\frac{-s}{1-\alpha}A^{1-\alpha}}}{1 + C \cdot e^{\frac{-s}{1-\alpha}A^{1-\alpha}}}, 0 \leq A < \infty, \\ N^*(0+) = 1 - N_0. \end{cases} \tag{12}$$

**Analysis and interpretation of the solution**

The function (8) can be written in the form:

$$\frac{dN}{dA} = L(A) \cdot P(A) \cdot E(A), \tag{13}$$

where

$$L(A) = \frac{C \cdot s}{(1 + Ce^{-\frac{s}{1-\alpha}A^{1-\alpha}})^2}, \quad P(A) = A^{-\alpha} \text{ and } E(A) = e^{-\frac{s}{1-\alpha}A^{1-\alpha}} \tag{14}$$

*L(A)* is the increasing bounded function varying from *Cs/(1+C)²* for *A = 0* until *Cs* for $A \to \infty$.

The product

$$P(A) \cdot E(A) = A^{-\alpha} \cdot e^{-\frac{s}{1-\alpha}A^{1-\alpha}} \tag{15}$$

is the power function with stretched exponential cut-off.

These two functions define two stages of a structure building. In the first stage, the elements (e.g. small fractures in the Earth's crust, microcracks in rocks and metals, etc.) grow in number and size independently because of their relatively low density in a substance. As the same mechanism is applied to each one of them, this process is scale invariant and is described by the power function $P(A) = A^{-\alpha}$ (Laherr`ere, Sornette 1998). The second stage is interpreted conceptually as interaction of elements, their coalescence and aggregation. This process is inverse to the multiplicative process described by stretched exponential function (Sornette, 1998; Naumis, Cocho, 2007). As these two processes are formally undistinguishable, same formula is applied to the second stage. Thus, one solution of generalized logistic equation models the process from birth and growth of elements to their coalescence and formation the whole structure. Figure 1 shows graphs of the function

$$\log_{10} N^*(A) = \log_{10} \frac{C \cdot e^{\frac{-s}{\alpha+1}A^{1-\alpha}}}{1 + C \cdot e^{\frac{-s}{\alpha+1}A^{1-\alpha}}} \ , \ C = 1000, \ s = 2.3, \ \alpha = 0.1 \qquad (16)$$

altogether with data from regions of Sumatra 2004 and Napa Valley 2014 earthquakes.

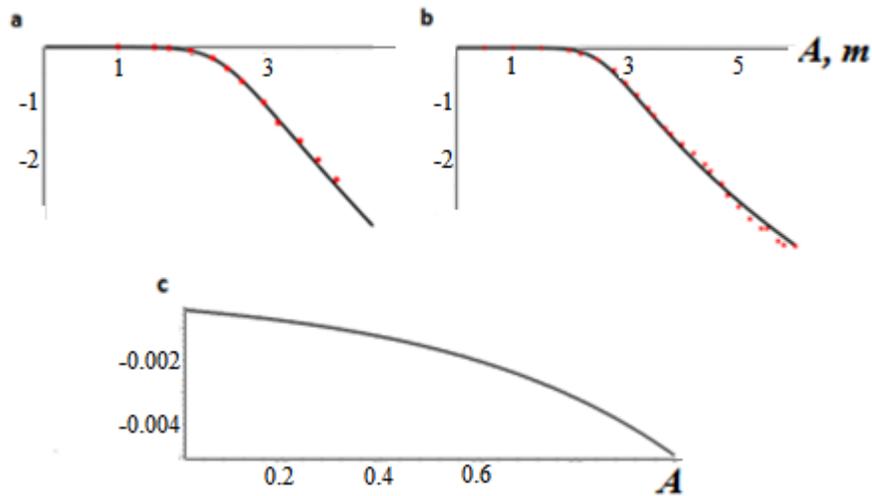

Figure 1

a. Theoretical function $\log_{10} N^*(A)$ and real data $\log_{10} N^*(m)$ from Sumatra 2004 earthquake region; b. Theoretical function $\log_{10} N^*(A)$ and real data $\log_{10} N^*(m)$ from Napa Valley 2014 earthquake region; $N^*(m)$ number of earthquakes with magnitudes $M \geq m$. Different intervals of earthquake magnitudes, time and depth intervals are tested; c. The "roll-off" part of $\log_{10} N^*(A)$ for $A \leq 1$.

The parameters $C$, $s$, and $\alpha$ in (7) for theoretical curves in Figures 1a, and b are found by solving the system of three equations (7) with points $(A_i, N_i)$, $i$=1,2,3, taken arbitrary from the observed data. The "roll-off" part of theoretical graphs and of the real data plots displayed on Figure 1 is not due to lack of data observed as it is stated in a number of publications, e.g. (Christensen, et al., 2002). It is the innate property of hierarchical aggregation processes. This part of a curve reflects the independent growth of elements in number and size, and because of their relatively low density in a substance it is described by the power function in (8) and (15). Earthquakes is one of examples of processes of that kind.

**Interpretation of the Gutenberg-Richter formula**

In this section, we substitute a variable $A$ in formula (12) by traditional in seismology letter $m$, magnitude of earthquake. The empirical G-R formula is:

$$\log_{10} N^*(m) = a - b \cdot m \tag{17}$$

$N^*$ is the number of all earthquakes of magnitudes $M \geq m$; $a$ and $b$ are constants. Using an expansion of (12) in powers of $C \cdot \exp(-s/(1-\alpha) \cdot m^{1-\alpha})$ for large $m$, we obtain

$$N^*(m) \approx C \cdot e^{\frac{-s}{1-\alpha} m^{1-\alpha}}, \tag{18}$$

and

$$\log_{10} N^*(m) \approx \log_{10} C - \frac{s}{1-\alpha} m^{1-\alpha} \cdot \log_{10} e \quad . \tag{19}$$

For $a = \log_{10} C, \alpha = 0, s = b$ we arrive at formula (17). Expression (19) is the analogy of

Gutenberg-Richter empirical formula with *b-value*

$$b = \frac{s}{1-\alpha} \cdot \log_{10} e . \tag{20}$$

In (19) $\alpha$ defines the concavity of the part of the $\log_{10} N^*(m)$ curve dominated by the stretched exponent *E(A)*. For $\alpha \approx 0$ this part of $\log_{10} N^*(m)$ is close to a straight line and interpreted in empirical data as a power dependence between magnitude *m* and total number of earthquakes with magnitudes $M > m$. Y. Malevergne and coauthors (Malevergne, Pisarenko, Sornette, 2005;

Malevergne, Sornette, 2006) in the course of statistical analysis of empirical data showed that the

power law family is asymptotically nested in the stretched exponential family. That is the

exponent *b* of the power law is obtained from the parameters of the stretched exponential fit

$\exp(-(m/d)^{\alpha})$ when $\alpha \to 0$. In the present work this result follows directly from solution (12)

for large *m* and $\alpha \to 0$.

**Relation with Tsallis non-extensive statistics**

For the unconstrained growth of a number of elements:

$$\left. \frac{dN}{dA} \right|_{gain} = s \cdot N^q \cdot A^{-\alpha}, \quad q > 0. \tag{21}$$

In Tsallis non-extensive statistical physics parameter *q* is interpreted as an "entropic index",

corresponding if *q*=1 to Boltzmann-Gibbs distribution (Vallianatos, et al., 2015).

For *0<q<1* the general solution of (21) is

$$N(A) = \left[ C + (1-q) \cdot \frac{s}{1-\alpha} \cdot A^{1-\alpha} \right]^{\frac{1}{1-q}}, A \in (0, \infty), 0 \le \alpha < 1 \tag{22}$$

and $C$ is an arbitrary positive constant. If $q>1$, then the equation (22) is valid for

$$0 < A < \left( \frac{C}{q-1} \cdot \frac{1-\alpha}{s} \right)^{\frac{1}{1-\alpha}}.$$

If $q=1$ and $0 \le \alpha < 1$, the general solution of equation (21) is

$$N(A) = C \cdot e^{\frac{s}{1-\alpha} \cdot A^{1-\alpha}}. \tag{23}$$

For $q=1$ and $\alpha=1$ the solution of (21) is a power-law function

$$N(A) = C \cdot A^s. \tag{24}$$

If $0<q<1$, and $\alpha=1$, the general solution of equation (21) is

$$N(A) = \left[ C + s(1-q) \cdot \ln A \right]^{\frac{1}{1-q}}, \tag{25}$$

$A > e^{-\frac{C}{s(1-q)}}$, and if $q>1$, $0 < A < e^{\frac{C}{s(q-1)}}$.

In case of a constrained growth of a structure due to loss of elements in the coalescence and aggregation, the main equation has the form

$$\frac{dN(A)}{dA} = s \cdot (1 - N(A)) \cdot N^q(A) \cdot A^{-\alpha}, \ q > 0. \tag{26}$$

The general solution of (26) for $q \in \mathbb{Z}^+$, $0 \le \alpha < 1$ in implicit form is

$$A = \left[ \frac{1-\alpha}{s} \cdot \left( \ln \frac{N(A)}{1-N(A)} - \sum_{n=1}^{q-1} \frac{1}{n \cdot N(A)^n} + C_1 \right) \right]^{\frac{1}{1-\alpha}} \tag{27}$$

provided the base to exponent $1/(1-\alpha)$ is not negative. Solution of the same equation for $0 < q < 1$, $\alpha \ne 1$ is

$$A = \left[ \frac{1-\alpha}{s} \cdot \sum_{n=1}^{\infty} \frac{N^{n-q}(A)}{n-q} + C_2 \right]^{\frac{1}{1-\alpha}} \qquad (28)$$

provided the base to exponent $1/(1-\alpha)$ is not negative. Figure 2 shows graphs of functions $N(A)$

from (27) and (28) for $C_1$=30, $C_2$=0.1, $\alpha$=0.1, $s$=2.5, $q = \frac{1}{2}$, $q$=1, and $q$=2. These graphs are plotted by reflecting graphs $A=N(A)$ about the line $N(A)=A$.

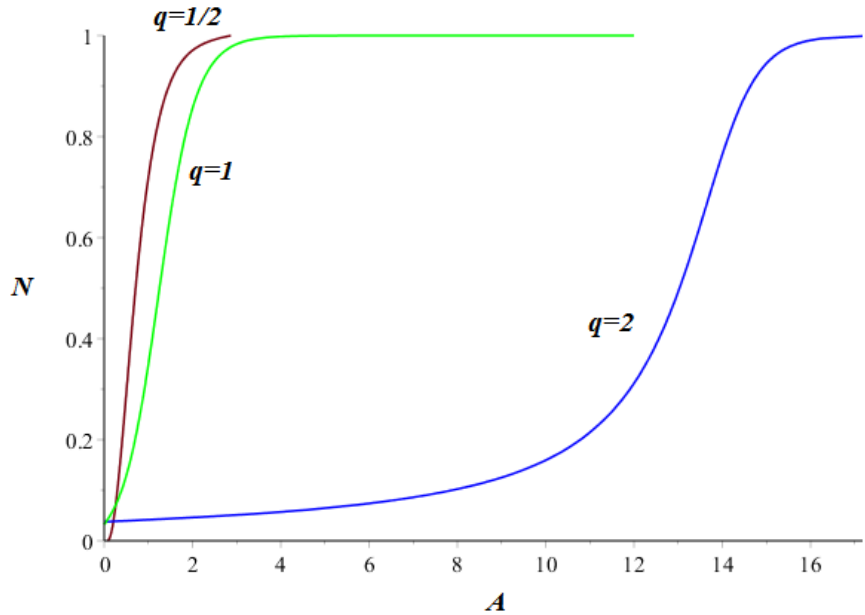

Figure 2

Solution of (28) for $q$=1/2 and solutions of (27) for $q$=1, $q$=2.

These graphs demonstrate distinction of aggregation processes in structures with different

properties. If to take (8) as a reference case corresponding to Boltzmann-Gibbs extensive

statistics, $q$=1, then solution of (28) represents the "super-additive" aggregation with the entropy

of a new system greater than entropies of components. Correspondingly, the solution of (27)

represents the "sub-additive" aggregation with the entropy of a new system less than entropies of components before the aggregation (Vallianatos, et al., 2015).

**Discussion and Conclusion**

The generalized logistic equation to model kinetics of natural processes such as earthquakes, forest fires, floods, landslides, and many others belonging to the class of hierarchical aggregation processes is derived. The solution of this equation is a product of an increasing bounded function and power-law function with stretched exponential cut-off. A. Clauset and coauthors (2009), using maximum likelihood fitting methods, tested twenty-four real world data sets including earthquake data. They demonstrated that the power-law with stretched exponential cut-off is "clearly favored over the pure power-law". Solution of generalized logistic equation confirms results of statistical analysis made by A. Clauset and coauthors (2009) and gives physics-based qualitative and quantitative approaches to the modeling of the hierarchical aggregation structures and processes. From solution (12) of this equation if follows that for large $m$ and $\alpha \rightarrow 0$ it asymptotically approaches the power function. This confirms result made in the course of statistical analysis of empirical data (Malevergne, Pisarenko, Sornette, 2005; Malevergne, Sornette, 2006). The exponent $\alpha$ defines the concavity of the part of the $\log_{10} N^*(m)$ real data curve dominated by the stretched exponent. See example on Figure 1b. Gutenberg-Richter formula results from interpretation of empirical data as a straight line in the area of stretched exponent with small $\alpha$. The solution of generalized logistic equation is applied for modeling distribution of foreshocks and aftershocks in the regions of Napa Valley 2014, and Sumatra 2004 earthquakes fitting the observed data well qualitatively and quantitatively. The relation of the theory developed with the Tsallis non-extensive statistics is demonstrated by solving the generalized logistic equation for $q>0$. It is shown that the case $0<q<1$ models "super-additive", and the case $q>1$ models "sub-additive" structures and processes. In the case $q=1$ the probability $p_{BG}$ of distribution of a number of elements in a structure corresponds to Boltzmann-Gibbs distribution. For $0<q<1$ and $0<N<1$ the ratio $N^{q-1}$ of probability of distribution of a number of elements to $p_{BG}$ is greater than 1, and the rate of change of number of elements with size is greater than the rate of "growth" of "Boltzmann-Gibbs structure", see for illustration graph for $q=\frac{1}{2}$ in the Figure 2. Also, for q>1 the ratio $N^{q-1}$ is less than 1 which results in a slower rate of change of number of elements with size, compared to that rate of in a "Boltzmann-Gibbs structure", graph for $q=2$ in the Figure 2.

**Data analysis**

To ensure the quality of modeling earthquake statistics, we use data from different data centers operating different networks of seismic stations: http://earthquake.usgs.gov/earthquakes/search/ (last accessed July 2015), and http://www.iris.edu/ieb/index.htm (last accessed July 2015). Data analysis is made by the variation of magnitude and depth ranges of earthquakes in selected regions.